\documentclass{aa}
\usepackage[pdftex]{graphicx} 
\usepackage{amsmath}
\usepackage{times}
\usepackage{natbib} 
\bibpunct{(}{)}{;}{a}{}{,} 

\newcommand{\gsim}{\,\raisebox{0.2em}{$>$}\!\!\!\!\!
\raisebox{-0.25em}{$\sim$}\,}
\newcommand{\lsim}{\,\raisebox{0.2em}{$<$}\!\!\!\!\!
\raisebox{-0.25em}{$\sim$}\,}
\newcommand{\kms} {km s$^{-1}$}
\newcommand{\lfir} {L$_{\rm FIR}$}
\newcommand{\msun} {M$_{\odot}$}

\newcommand{\urad}{U_{\rm rad}}
\newcommand{\ub}{U_{\rm B}}  

\newcommand{\tauloss}{\tau_{\rm loss}}

\newcommand{\vshock}{v_{\rm shock}}

 \newcommand{\gr}{$\gamma$-ray \,}
 \newcommand{\grs}{$\gamma$-rays \,}

\begin{document}
 \title{Shock acceleration of relativistic particles in galaxy-galaxy
 collisions}


  \author{U. Lisenfeld\inst{1,2}
          \and
          H. V\"olk\inst{3}          }

       \institute{Departamento de F\'\i sica Te\'orica y del Cosmos,
         Universidad de Granada, 18071 Granada, Spain, \email{ute@ugr.es} \and
         Instituto `Carlos I' de F'sica Te—rica y Computacional, Universidad de Granada, Spain \and
          Max-Planck-Institut f\"ur Kernphysik, Postfach 103980, 69029 Heidelberg, Germany,
          \email{Heinrich.Voelk@mpi-hd.mpg.de} 
            }
  \date{   }

  \abstract 
{All galaxies without a radio-loud AGN follow a tight correlation between their
  global FIR and radio synchrotron luminosities, which is believed to be
  ultimately the result of the formation of massive stars. Two colliding pairs
  of galaxies, UGC12914/5 and UGC 813/6 deviate from this correlation and show
  an excess of radio emission which in both cases originates to a large extent
  in a gas bridge connecting the two galactic disks. } 
{We are aiming to clarify the origin of the radio continuum
  emission from the bridge. } 
{The radio
  synchrotron emission expected from the bridge regions is calculated, assuming
  that the kinetic energy liberated in the predominantly gas dynamic
  interaction of the respective interstellar media (ISM) has produced shock
  waves that efficiently accelerate nuclei and electrons to relativistic
  energies. We present a model for the acceleration of relativistic
  particles in these shocks and calculate the resulting
  radio emission, its spectral index and the expected high-energy \gr
  emission. } 
{ It is found that the nonthermal
  energy produced in the collision is large enough to explain the radio
  emission from the bridge between the two galaxies. The calculated spectral
  index at the present time also agrees with the observed value. The
  expected \gr emission is not detectable with present day instruments, but
  might be observable with the next generation of \gr telescopes. } 
{The deviation of
  these two interacting galaxy systems from the standard FIR-radio correlation
  is consistent with the acceleration of an additional population of electrons
  in large-scale shock waves resulting from the gas dynamic interaction of the
  two ISM.
    This process is not related to star formation and therefore it is expected
    that the systems do not follow the FIR-radio correlation. In particular,
    the radio emissions of these systems do not represent an argument against
    the calorimeter theory.  The acceleration of relativistic electrons in
    shocks caused by an ISM collision, in the same way as described here, is
    likely to take place in other systems as well, as in galaxy clusters and
    groups or high-redshift systems.  } 
\keywords{cosmic rays -- Galaxies: interaction -- Galaxies: ISM -- Galaxies:
    invididual (UGC 12914/5, UGC813/6) -- Radio continuum: galaxies }

 \maketitle

%

\section{Introduction}
Since the early years of the IRAS mission in far-infrared (FIR) astronomy, late
type galaxies without a bright, radio-loud AGN are known to show a universal,
tight correlation between their spatially integrated FIR luminosities and
monochromatic radio continuum emissions  \citep{dejong85,helou85}. It is
followed by all galaxies later than SO/a regardless of morphology, size, color,
etc. For reviews, see e.g. \citet{condon92,voexu94}. The correlation is
believed to be ultimately a result of the formation of massive stars. These
produce the optical/UV radiation which heats interstellar dust grains. The FIR
emission is then the thermal radiation of those heated grains. At any given
time some of the stars have a strong mass loss, either due to winds or due to
their ultimate explosion as supernovae, which may lead to the acceleration of
electrons to relativistic energies and allows them to emit radio synchrotron
radiation.

The identification of the dominant physical processes and the quantitative
determination of their role in galaxies has been a matter of discussion ever
since the correlation was found. One explanation is the calorimeter theory
\citep{voelk89}. It argues that the main fraction of the dust heating stellar
photon flux is absorbed by the dust grains in the galaxy, and that the
interstellar relativistic electrons dominantly lose their energy within the
galaxy and its halo by Inverse Compton collisions with the photons of the
interstellar radiation field, as well as by synchrotron radiation in the magnetic
field, rather than escaping from the galaxy without major energy loss, like the
nuclear cosmic ray particles. Thus galaxies act like calorimeters for their own
primary photon and electron emissions, independent of the details of production
and transport of these carriers of energy. 

The theory has been disputed on the grounds that the synchrotron frequency
spectrum of a significant fraction of galaxies does not show the steepness
expected in a simple one-zone transport model of galactic escape by spatial
diffusion in the presence of radiative energy losses \citep[e.g.][]{niklas97}.
It has also been demonstrated that in a variety of cases the power output in
the UV is comparable to that observed in the FIR -- defined in different ways
\citep{popescu02,xu06} -- showing that not all normal galaxies are optically
thick to their own UV emission.  However, \citet{lisenfeld96} showed that the
calorimeter theory is likely to hold in an {\it approximate} way for normal
spiral galaxies.  Furthermore, the spectral index of the synchrotron frequency
spectrum of an entire galaxy does not allow a precise estimate of the escape
fraction. Rather the spatial variation of the spectral index in the halo should
be used \citep{lisenfeld00}. 
{  Most recently, \citet{strong10} have argued, using the GALPROP model for cosmic ray propagation, that  our Galaxy is indeed nearly a cosmic ray electron calorimeter.}
A recent discussion of the theoretical basis for
the FIR-radio correlation is contained in \citet{lacki10}.

A key element for the understanding of the correlation is the question whether
the synchrotron electrons escape from galaxies without essential energy loss or
not. In this context, \citet{condon93,condon02} have investigated two face-on
interacting spiral galaxy systems where, presumably after the interaction, the
respective pairs of galaxy disks are well separated from each other optically,
but are connected by a radio continuum-bright ``bridge'' of gas, suggested to
be stripped from the interpenetrating disks. \citet{condon93,condon02}
gave an interpretation of the fact that the two systems
show overall a significant excess of radio continuum emission relative to the
FIR-radio continuum ratio, expected from the FIR-radio correlation for single
galaxies. They concluded that the excess radio emission in the so-called
bridges was rather the result of the escape of relativistic electrons from the
respective gas disks and that it was only visible because
the stretched and therefore strong magnetic field in the bridges allowed them
to emit synchrotron radiation. The authors contended that in noninteracting
galaxies a similar escape would occur,  albeit silently,
because of the negligible magnetic field strength in their outer
regions. Therefore they concluded that isolated disk galaxies were not
good ``calorimeters''.

In this paper the dynamical effects of such galaxy-galaxy collisions on the
interstellar gas are investigated. It is argued that the interstellar media of
the respective galaxies will undergo a largely gas dynamic interaction, where
the low-density parts exchange momentum and energy through the formation of
large-scale shock waves in the supersonic collision. The stars and the dense
parts of the more massive interstellar clouds, on the other hand, will
interpenetrate each other with only a stellar-dynamic interaction. To lowest
approximation the stellar disks will asymptotically remain intact, together
with the massive interstellar clouds, even though the latter may be stripped of
their outer layers. The lower-density interstellar media, on the other hand,
bring each other to rest in their center of momentum system, and remain between
the seperating stellar disks. The interstellar magnetic fields, anchored
primarily in the massive molecular clouds, are thereby stretched and their
tension substantially increased. It is then argued that the resulting
interstellar shocks should be able to diffusively accelerate charged nuclear
particles and electrons to ultrarelativistic energies in a very effective
manner, comparable to what typically happens in supernova remnants.  This leads
to the large-scale production of a new population of relativistic electrons,
diffusively confined in the fluctuating magnetic fields -- which themselves are
possibly also amplified by the accelerating particles, even though this effect
is neglected here. The new electron population then implies significant
synchrotron emission from the interaction region. In this paper a model for the
acceleration of relativistic particles is presented and the synchrotron
emission from the relativistic electron component is calculated, as well as the
expected \gr emission from relativistic nuclei and electrons. This model is
shown to be able to explain the radio continuum emission observed from the
bridge between the galaxies.

\section{The two collisional galaxy systems}

\begin{table*}
      \caption[]{Radio and FIR properties of the two systems.}
         \label{data1}
\begin{tabular}{lcccc} 
  \hline
  & UGC 12914/5      & Ref.      & UGC 813/6     & Ref.   \\ 
  \hline\hline
  Distance                                         & 60 Mpc &    & 69 Mpc  & \\
  S$_{\nu}$- system &   114 mJy & 1&62.9 mJy &  2 \\
  S$_{\nu}$ - individual &  19mJy (UGC 12914) &1  &  19mJy (UGC 813) & 2  \\
  & 47mJy  (UGC 12915)  &1 &  22mJy (UGC 816) & 2 \\
  P$_{\nu}$ - system & 4.9\,10$^{22}$ WHz$^{-1}$ && 3.6\,10$^{22}$ WHz$^{-1}$ &  \\
  P$_{\nu}$ - bridge & $(1.7-2.1) \times10^{22}$ WHz$^{-1}$ & 1& $(1-1.2) \times10^{22}$ WHz$^{-1}$ & 2 \\
  $\alpha$ - bridge & 1.3-1.4 & 1& 1.3 & 2\\
  S$_{60\mathrm{\mu m}}$ - system & 6.3 Jy & 3 &2.76 Jy & 4 \\
  S$_{100\mathrm{\mu m}}$ - system & 13.4 Jy & 3 &7.58 Jy & 4\\
  \lfir   -system  &  1.6 \,10$^{37}$ W &  & 1.05 \,10$^{37}$ W & \\
  $q$             & 1.93  &&1.90 & \\
  \hline
\end{tabular}

See Sect. 2 for a detailed description of the entries.

References:  (1) \citet{condon93} (2) \citet{condon02} (3) \citet{jarrett99}  (4) \citet{bushouse88}

\end{table*}


\citet{condon93}  described the radio continuum and HI properties of UGC
12914/5, called "Taffy" galaxies, a system of two galaxies that have
experienced a direct collision some 30 Myr ago, in which the galactic discs
collided head-on and interpenetrated each other.  A bridge
 of synchrotron emission extends between both stellar disks,
showing that both relativistic electrons and magnetic fields 
are present. The spectral index of the radio emission between 1.49 and
4.9 GHz steepens gradually from the stellar disks with values of 0.7--0.8 to
values of 1.3--1.4 in the middle of the connecting gas bridge. This steep
spectral index will be argued to be indicative of dominant synchrotron and
inverse Compton losses suffered by the relativistic electrons.

A further noticeable feature of this system is that the ratio between the FIR
and the radio synchrotron luminosities is about a factor of 2 lower than the
mean value found in spiral galaxies.  One can describe this ratio by the
parameter $q$ \citep{helou85}:

\begin{equation}
q = \log \left(\frac{FIR}{3.75 \times 10^{12} {\rm W m^{-2}}}\right)-
\log \left(\frac{S_{\rm 1.4 GHz}}{\rm W m^{-2}Hz^{-1}}\right),
\end{equation}
where $S_{\rm 1.4 GHz}$ is the observed 1.4 GHz flux density in units of 
W m$^{-2}$Hz$^{-1}$ and where

\begin{equation}
FIR = 1.26 \times 10^{-14}(2.58 S_\mathrm{60\mu m} + S_\mathrm{100\mu m}),
\end{equation}
with $ S_{\rm 60\mu m}$ and $S_{\rm 100\mu m}$ denoting the IRAS 60 and 100
$\mu$m band flux densities in units of Jy.  The value for the galaxy system is
$q=1.93$, lower by a factor of about two than the values found for samples of
moderately luminous FIR galaxies without an active nucleus. For example, Yun et
al. (2001) derived a mean value of $q=2.34$ with a mean deviation of $0.25$ for
a sample of 1809 galaxies with $S_\mathrm{60\mu m} \ge 2$ Jy.  Since about 40\% of the
total radio continuum emission comes from the bridge, but 
only a much smaller fraction of the total dust emission ($<$ 5 \% at 15
$\mu$m, \citealt{jarrett99}, and about 20\% at 450 and 840 $\mu$m,
\citealt{zhu07}), the reason for this low value of $q$ is an excess radio
emission from the bridge.

Apart from the sychrotron emitting electrons the bridge contains also large
amounts of atomic (Condon et al. 1993) and molecular (Braine et al. 2003) gas.
The total amount of gas in the bridge is several times $10^9$ \msun.  An
efficient way of depositing these gas masses in the bridge is a gas dynamic
interaction of the diffuse parts of the galaxies' interstellar media (ISM).  It
is expected that also the extended HI clouds collide efficiently in a direct
collision.  Braine et al. (2004) argued that this mechanism is even valid for
the smaller and denser molecular gas clouds traced by CO. The gas gets ionized
in this high-speed collision but, due to the high densities, it recombines and
becomes molecular again on a very short time-scale. The line ratios and the
isotopic ratios of CO show that the opacities are low, and no high-density
tracers such as HCN have been detected. Our interpretation is that most likely the
dense cores -- capable of forming stars -- have remained within the stellar
disks together with the stars. Zhu et al. (2007) showed that the physical
conditions of the molecular gas in the bridge are comparable to those in the
diffuse clouds in our Galaxy. In agreement with this interpretation, the SFR in
the bridge is only about 10\% of what is expected from the molecular gas mass
\citep{braine04}.  Also, the mid/FIR emission from dust in the bridge is very
low \citep{jarrett99,zhu07}, indicating a very low SFR, but showing that some
dust is nevertheless present.

A similar system, UGC 813/6 was described in a later paper by Condon et
al. (2002).  Features common to both systems are (i) a synchrotron bridge
between the two disks (ii) a low value of the parameter $q$, and (iii) the
presence of large amounts of atomic and molecular gas in the bridge. The
similarities indicate that this second system has gone through a similar
collision process, and that the gas has
 been deposited by the same mechanism in the bridge.

 In Tab.~\ref{data1}  some observational data for these systems are listed.
 The entries in the table are:
 
 \begin{description}
 \item[Row 1:] Adopted distance, based on $H_0 = 75$ \kms Mpc$^{-1}$.
 \item[Row 2:] Total radio flux density of the systems obtained from VLA
   observation in the D-array (UGC 12194/5), respectively in the C-array (UGC
   813/6).  The frequency of the radio observations is 1.49 GHz (UGC 12194/5),
   respectively 1.40 GHz (UGC 813/6)
 \item[Row 3 and 4:]   Radio flux density of the individual galaxies  from
 VLA B-array observations.
 \item[Row 5:]  Radio continuum luminosity of the systems obtained from the flux density in row 2.
\item[Row 6:]  Radio continuum luminosity from the bridge. The lower value is calculated
from the difference between the value for the system and the individual galaxies in
the B-array observations. The higher values uses the slightly higher
C (respectively D) array values for the
emission from the total system and thereby assumes that the difference between the higher and lower resolution observations  is due
to diffuse emission coming from the bridge.
\item[Row 7:] The radio spectral index between 1.49 and 4.86 GHz (for UGC12914/5)
and between 1.40 and 4.86 (for UGC 813/6)  in the middle of the bridge 
\item[Row 8 and 9:] The IRAS flux densities at 60 and 100 $\mu$m.
 \item[Row 10:] The FIR luminosity calculated as:   \\
    \lfir $=1.5 \times 10^{32} \left(\frac{D}{\rm Mpc}\right)^2 \left(2.58 S_{\rm 60\mu m} +
S_{\rm 100\mu m}\right) [W]$
 \item[Row 11:] The radio-to-FIR ratio, $q$, calculated from eqs. (1) and (2).
 
 \end{description}

\section{Modelling the radio emission from the bridge}

\subsection{Energy released in  the interaction of the ISM}

In a face-on collision, as suffered by UGC 12194/5 and UGC 813/6, the stellar
disks interpenetrate each other without being too much altered. However, the
diffuse gas and part of the gas clouds interact hydrodynamically and exchange
energy and momentum. The large quantities of atomic and molecular gas that are
present in both brigdes clearly show that such gas interaction took place
(Condon 1993, 2002; Braine 2003, 2004), even though the denser parts of the gas
clouds should survive like the stars.

If one assumes that half of the gas, which is now
present in the bridge, was previously in one galaxy, and the other half in the
other galaxy,  then the total energy liberated in a fully inelastic
interaction is the kinetic energy of the gas mass:

\begin{equation}
E_{\mathrm kin}  = \frac{1}{2}  M_\mathrm{gas} \left(\frac{v_\mathrm{coll}}{2}\right)^2,
\end{equation}
where $M_\mathrm{gas}$ is the total gas mass in the bridge and
$v_\mathrm{coll}$ denotes the velocity difference of the gas at
collision. The factor 1/2 converts this velocity to the velocity difference in
the center of mass system (assuming that both gas disks are equally massive).

With respect to the gas masses, for UGC 12914/5 the values given
in Tab. 1 in \citet{braine03} are used.  For UGC 813/6 the masses from the
bridge were calculated from the values of the three diagonal pointings (at
offsets (0,0), (-7,10) and (7,10) given in Tab. 1 of \citet{braine04}).  For
the determination of the molecular gas mass from the CO measurement, a
conversion factor has to be adopted. Braine et al. (2003) argued that the
conversion factor in the bridge of these interacting systems is most likely
lower than the Galactic value of $N(H_2)/I_{CO} = 2 \times 10^{20}$ cm$^{-2}$
(K \kms)$^{-2}$, due to the low opacity indicated by the isotopic line ratios
and the CO/HCN ratio.  They suggest that a four times lower conversion factor
is a good estimate, in agreement with the results of \citet{zhu07}, based on a
multi-transition analysis.  This is indeed possible if the molecular gas in the
bridge region is of rather low density, having been stripped from denser
molecular clouds. Here, both values are adopted, the Galactic conversion factor
and a 4 times lower factor, in order to cover the range of likely values.

The relative velocity between the galaxies, $v_{\rm coll}$, in the case of UGC
12914/5 has been derived by \citet{condon93} from a dynamical analysis of the
HI line and of the galaxy masses. In addition to the relative velocity, the
disks are counterrotating at a velocity of about 570 \kms .  However, since the
collision is face-on, this counterrotation will only make this an oblique
shock.  Given the expected smallness of the Alfv\'en Mach number, the velocity
shear will have a negligible effect on the shock compression and the resulting
particle acceleration.  Therefore this shear velocity component is not taken
into account in the calculation of the available energy for particle
acceleration.  In the case of UGC 813/6, $v_{\rm coll}$ is estimated following
the analysis of \citet{condon02} in their Sect. 3.4, but correcting their
eq. (1) to yield:

\begin{equation}
\left(1+\frac{v_\parallel^2}{v_\perp^2}\right)^{\frac{1}{2}}
\left(1+\frac{v_\perp^2}{v_\parallel^2}\right) =
\frac{2 G M_{\rm T}}{\delta_\perp v_\parallel^2},
\end{equation}
where $v_\parallel$ ($v_\perp$) is the velocity component parallel
(perpendicular) to the line of sight, $\delta_\perp$ is the distance between
the galaxies perpendicular to the line of sight, $G$ is the gravitational
constant and $M_{\rm T}$ the total mass of the system. Using the numbers derived for
$M_T$, $v_\parallel$ and $\delta_\perp$ in \citet{condon02}, $v_{\rm coll} =
\sqrt{v_\parallel^2+v_\perp^2} \approx 600$ \kms , sligthly higher than the
value of 500 \kms\ derived by \citet{condon02}.

The masses in the bridge,  the collision velocities and the resulting kinetic energies
for both systems are listed in Tab.~\ref{data2}.
The gas masses are multiplied by a factor
of 1.37 to take into account the helium fraction.  
  

\begin{table*}
      \caption[]{Modelling the radio emission}
         \label{data2}
\begin{tabular}{lcccc} 
\hline
                             & UGC 12914/5      & Ref.     & UGC 813/6     & Ref.   \\ 
 \hline\hline
Kinetic energy  & & & & \\
\hline
$v_{\rm coll}$   & 600 \kms                             & 1  & 600 \kms& 2, 5 \\
 M(HI)$^\mathrm{(a)}$(bridge) & $2\, 10^9$\msun & 3 & 3.2 \,10$^9$ \msun & 4\\
 M($H_2$)$^\mathrm{(a,b)}$(bridge) & $(2-9)\times10^9$ \msun& 3 & $(0.55-2.2) \times 10^9$\msun & 4\\
$E_\mathrm{kin,low}^{(c)}$  & $4.9 \,10^{57}$erg   &  5 &$3.2 \,10^{57}$erg& 5 \\
$E_\mathrm{kin,high}^{(c)}$  & $1.4 \,10^{58}$erg   &  5 &$4.6 \,10^{57}$erg& 5 \\
\hline
\multicolumn{2}{l}{Energy loss of the relativistic electrons} & & & \\
\hline
 distance between galaxy nuclei & 20 kpc  &  1& 21.6 kpc & 2 \\
time since start of interaction, $T$ & $2.8 \,10^7$ yr &5  &$3.1 \,10^7$ yr &  5 \\
$B$             & 7 $\mu$G  & 1 & 7$\mu$G & 2 \\
$U_\mathrm{rad}$ &  0.76 eV cm$^{-3}$  & 5& 0.56 eV cm$^{-3}$ & 5 \\
\hline
Predicted radio emission& & & & \\
\hline
P$_\nu$-predicted,low$^\mathrm{(c,d)}$ & $(0.6-1.8)\times10^{22}$ W Hz$^{-1}$  & 5 & $(0.6-1.8)\times 10^{22}$  W Hz$^{-1}$ & 5\\
P$_\nu$-predicted,high$^\mathrm{(c,d)}$ & $(1.7-5.1)\times10^{22}$ W Hz$^{-1}$  & 5 & $(0.9-2.6)\times 10^{22}$  W Hz$^{-1}$& 5\\
$\alpha$-predicted & 1.4 &5& 1.3 & 5 \\
\hline
\end{tabular}

See Sects. 3.1 and 3.4 for a detailed explanation how the quantities were derived.

$^{(a)}$  The masses include a factor of 1.37 for the Helium fraction.

$^{(b)}$  The higher mass is calculated for a Galactic conversion factor from
CO intensity to H$_2$ mass, the lower value for a 4 times  lower conversion factor.

$^{(c)}$ The value named "low" is calculated with a 4 times lower value for the molecular
gas mass 
and the value named ``high"  is calculated with a Galactic conversion factor.

$^{(d)}$ The range of values is given by the uncertainties of the efficiency
of CR acceleration in shocks (between 10 and 30\%)

References:  (1) Condon et al. (1993) (2) Condon et al. (2002)  (3) Braine et al. (2003) (4) 
Braine et al. 2004
 (5)  present work

\end{table*}

\subsection{Production of relativistic electrons}

During the collision of the ISM, strong large-scale shocks will form.  Since
the magnetosonic velocity in the ISM, of the order of some 10 \kms, is much
smaller than the velocity of the collision, of several 100 \kms, the square of
the magnetosonic Mach number of these shocks will be $M_\mathrm{ms}^2 = O(100)
\gg 1$. A  tangential discontinuity will form at the
position of the collision and two shocks will propagate in 
  opposite directions with velocities $v_\mathrm{shock}$, communicating
  the interaction to larger and larger fractions of the colliding interstellar
  gas masses. {  In between these shocks the galaxy is  filled with 
  post-shock gas. Support for this prediction comes from the
  observation of strong, shock-excited H$_2$ emission within the bridge of UGC 12914/5
 (Peterson et al., in preparation).}

  {  Fig. 1 shows the idealized picture of this interaction, the basis of our
  model,}
   in the reference frame of  the
  motion normal to  this  tangential discontinuity, which
is also the center of mass system. In this frame, the post-shock normal velocity
$v_\mathrm {post}$ vanishes and  the preshock normal velocity of the
gas  is $v_\mathrm {pre}=\frac{1}{2}v_\mathrm{coll}$. 
 {  The contact discontinuity is stationary and situated in the middle between
the galaxies (at $x=0$ in Fig. 1).}
 In a strong,
but approximately unmodified, adiabatic gas shock,
the normal component of the velocity difference between the shock in the
pre- and postshock gas follows the relation:
 
 \begin{equation}
   4 \times (v_\mathrm{post}-v_\mathrm{shock}) =
   (v_\mathrm{pre}-v_\mathrm{shock}).
 \end{equation}
 With $v_\mathrm{post}=0$, this yields
 $v_\mathrm{shock}=\frac{1}{6}v_\mathrm{coll}$.  The shock velocity
   $V_\mathrm{s}$, relative to the unperturbed ISM gas, is then $V_s =
   v_\mathrm{shock} + \frac{1}{2}v_\mathrm{coll} = \frac{2}{3} v_\mathrm{coll}$ = 400 km/sec,
   for UGC 12914.
   
 {  We can make a rough estimate of the present position of the shocks,
 based on the geometry of the galaxies. Taking UGC 12194/5 as an example,
 we estimate from Fig. 1 in 
 \citet{condon93} that the distance between the
 galaxy midplanes is roughly twice the thickness of the galaxy disks, $L$.
 We can then calculate the time $T$ elapsed since the beginning of the
 interaction between the ISM in the galaxy disks, taking into account
 their thickness $L$, 
  as $T= 1.5L/( \frac{1}{2}v_\mathrm{coll})$. The shocks that formed
 at this time have propagated since then  in opposite directions with a speed
 $v_\mathrm{shock}$, reaching by now a distance from the
 contact discontinuity (at $x=0$) of $v_\mathrm{shock} T = 
 \frac{1}{6}v_\mathrm{coll} T = L/2 $. 
 Thus, in UGC 12194/5 the shocks are expected to be very close to the
 galaxy disks so that practically the entire bridge is expected to be
 filled with post-shock gas.}

 The Mach number of the shocks produced in this collision is like in
 middle-aged supernova remnants in the Sedov phase. Roughly speaking, the
 particle acceleration efficiency of such shocks will therefore be
 similar to that of a supernova remnant, i.e. of the order of $10 - 30 \%$
 \citep[e.g.][]{berezhko97}.  This is a basic assumption for the present
   paper.

 The source function $Q(E)$, i.e. the number of relativistic particles produced
 by the shock per energy and time interval is given by:

 \begin{equation}
Q (E) = 2 f_\mathrm{acc}(E) v_\mathrm{shock} A
\end{equation}
 where  $A$ is the area covered by the shock (roughly the area of the galaxy disks).
The factor 2 is due to the fact that the shocks propagate into two, opposite directions.
$ f_\mathrm{acc}(E)$ is the downstream, uniform number of relativistic
particles of rest mass $m$, produced per volume and energy interval 
 at  the  shocks:

 \begin{equation}
f_\mathrm{acc}(E) = f_0  \biggl({E\over m c^2}\biggr)^{-\gamma} .
\end{equation}
Here, $\gamma$ is the spectral index of the differential relativistic particle
source spectrum, taken to be $\gamma=2.1$ \citep{drury94,berezhko97}.  For
  the nuclear particles, essentially protons, the constant $f_0$ can be
determined by requiring that the total energy 
  converted into relativistic particles during the entire duration of the
interaction, $T$, is equal to $E_\mathrm{acc}=(0.1-0.3) \times E_\mathrm{kin}$:
 
 \begin{eqnarray}
   E_\mathrm{acc} & = &  \int_{E_{\rm min}}^\infty E\, Q(E)\, T\, dE \\
   & = &  \int_{E_{\rm min}}^\infty 2\,E f_0  \biggl({E\over m_\mathrm{p}c^2}\biggr)^{-\gamma}  v_\mathrm{shock} 
   A\,  T\,  dE \nonumber \\
   & =  & 2 f_0 {(m_\mathrm{p}c^2)^2 \over \gamma-2}  \biggl({E_{\rm min}\over m_\mathrm{p}c^2}\biggr)^{-\gamma+2} 
   v_\mathrm{shock} A\,  T,
   \nonumber \\
\end{eqnarray} 
%
for $\gamma >2$. Here, $E_\mathrm{min}$ is the minimum energy of relativistic 
  protons of mass $m_\mathrm{p}$ accelerated, taking as
$E_\mathrm{min}= 1$ GeV \citep{drury94}.  This gives 

\begin{equation}
 f_0  = \frac{E_\mathrm{acc}}{ 2 T \, A\, v_\mathrm{shock} } 
 \frac{(\gamma-2)}  {(m_\mathrm{p}c^2)^2}
 \left(\frac{E_\mathrm{min}}{m_\mathrm{p}c^2}\right)^{\gamma-2}. 
\end{equation}

In the Galactic cosmic rays, at a given energy, the number of relativistic
electrons is about 1\% of that of the protons at GeV energies, and the source
spectra for electrons and protons are probably similar
\citep{Mueller01}. Assuming that this electron-to-proton ratio is also
representative for the source spectra in the galaxies considered here, implying
that the electron and proton source spectra are the same, we adopt for the
source function of the relativistic electrons $Q_{\rm e}(E) = Q(E) \times 0.01$
which means that their source distribution is $f_\mathrm{acc,e} =
f_{0,\mathrm{e}} (E/(\mathrm{m_\mathrm{p}} c^2))^{-\gamma}$, with
$f_{0,\mathrm{e}} = f_0 \times 0.01$.

\begin{figure}
\includegraphics[width=9.cm]{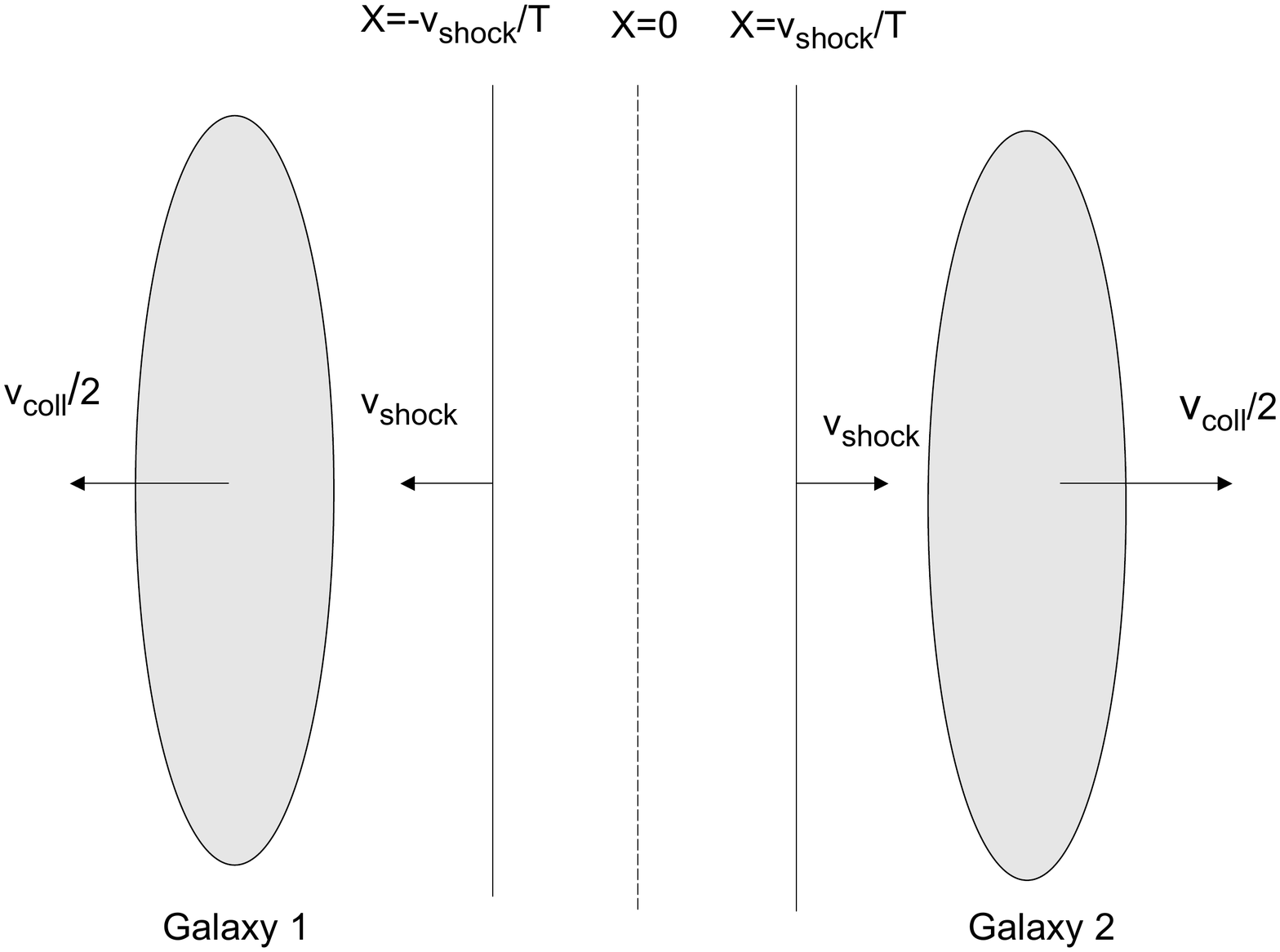}
\caption{Schematical illustration of the shock in the bridge region (see Sect. 3.2
for more explanations) {  in the center of mass system.}}
\end{figure}

\subsection{Relativistic electron density and  synchrotron emission}

In order to calculate the synchrotron emission from the bridge we solve the
 time-dependent equation for the electron particle density
 $f_\mathrm{e}(t,x,E)$. Due to the spatial symmetry of the situation, we can
 limit ourselves to a one-dimensional approximation where $x$ is the coordinate
 in the direction along which the galaxies separate (see Fig. 1).  Furthermore,
 we neglect the diffusion of relativistic electrons because the typical spatial
 scales which are relevant on the time-scales discussed here, $(3 -4)\times
 10^7$ yr, are $\lsim$ 1 kpc, whereas the width of the bridge is 10 kpc.
We can then write:

\begin{equation}
\label{prop_cre}
{\partial f_\mathrm{e}(t,x,E)\over \partial t}
        =  q_{\rm e}(t,x,E) +
    {\partial\over \partial E} \biggr\{b
     \, E^2 \, f_\mathrm{e}(t,x,E)\biggr\}.
\end{equation}
where 
\begin{equation}
  q_{\rm e}(t,x,E) = \frac{Q_{\rm e}}{A} \bigl(\delta(x-v_\mathrm{shock}t)+ \delta(x+v_\mathrm{shock}t)\bigr)
\label{source}
\end{equation}
is the local source strength (in units of relativistic electrons produced
per energy interval per time and per volume).  This source strength describes
two shocks that start at $t=0$ at $x=0$ and propagate
into  opposite directions with velocity $v_\mathrm{shock}$.  Eq.\ref{prop_cre}
takes into account the electron acceleration in the shocks and the subsequent
radiative energy losses of these CR electrons due to Inverse Compton and
synchrotron losses as:
\begin{eqnarray}
\left(\frac{\mathrm{d}E}{\mathrm{d}t}\right)_\mathrm{rad} & = & -bE^2 = -
\frac{4}{3} \sigma_\mathrm{T} c \left(\frac{E}{m_\mathrm{e}c^2}\right)^2
(\ub+\urad)   \nonumber \\
\end{eqnarray}
where $\sigma_\mathrm{T}$ is approximated by the Thompson scattering cross
section, $B$ the magnetic field strength, $\ub$ its energy density and $\urad$
is the energy density of the radiation field.
From the radio and optical data presented by Condon et al. (1993) one can
conclude that at the present time the two stellar disks are seperated by a
distance that is roughly equal to the sum of their optical thicknesses. This
implies that the shocked material in the bridge between them has barely
expanded back to its preshock density. Therefore the adiabatic losses of the
accelerated electrons can probably be disregarded relative to the radiative
losses. This is consistent with the fact that the distribution of the
synchrotron spectral index $\alpha(1.49,4.86)$, between 1.49 and 4.86 GHz, has
an integrated value $\alpha \gsim 1.00$ in the bridge region (Condon et al.,
1993), indicative of dominant Inverse Compton and synchrotron losses.

The solution of eq.\ref{prop_cre} is simplyfied by noting the symmetry
of the local source function (eq.\ref{source}) around $x=0$, so that
$f_\mathrm{e}(E,x,t) = f_\mathrm{e}(E,-x,t)$. 
The solution of eq.~\ref{prop_cre}  is:

\begin{eqnarray}
  f_\mathrm{e}(E,|x|,t)  &  &=  f_{\rm 0,e}
  \biggl({E\over \mathrm{m}c^2}\biggr)^{-\gamma}
  \left\{ 1-E\,b \left(t-\frac{|x|}{v_\mathrm {shock}}\right)
  \right\}^{\gamma-2} \nonumber \\ & & \text{for $t > |x|/\vshock$ and
  $t-|x|/\vshock < \tauloss$} \nonumber \\ f_\mathrm{e}(E,|x|,t) & & = 0
  \hspace{2.cm} \nonumber \\ & & \text{for $t \le |x|/\vshock$ and
  $t-|x|/\vshock \ge \tauloss$}
  \label{dist_f}
\end{eqnarray}
with $\tau_\mathrm{loss} = (Eb)^{-1}$ being the life-time of a relativistic
electron against radiative energy losses.  This means that the density of
relativistic electrons, $f_\mathrm{e}(E,x,t)$, is zero in those regions, where
either the shock has not passed yet (for $t \le |x|/v_\mathrm{shock}$ ) or
where all relativistic electrons have already lost their energy down to levels
below $E$ (for $t - |x|/v_\mathrm{shock}\ge \tau_\mathrm{loss} $).

We integrate this expression over the volume of the bridge and obtain the total
number of relativistic electrons in the bridge at time $t$ and per energy
interval, $F_\mathrm{e}(E,t)$. Due to the symmetry of $f_\mathrm{e}(E,x,t)$
with respect to $x=0$ we can carry out this integration only for $x\ge 0$ and
multiply the result by a factor of 2:

\begin{equation}
F_\mathrm{e}(E,t) = 2\, A \int_{x_\mathrm{min}}^{x_\mathrm{max}}
f_\mathrm{e}(E,x,t) dx
\end{equation}
where 
$x_\mathrm{min} = \max\left\{0,(t- \tau_\mathrm{loss})v_\mathrm{shock}\right\}$ 
and $x_\mathrm{max} = t\, v_\mathrm{shock}$ because outside
these limits $f(E,x,t) = 0$.  The result of the integration is:

\begin{eqnarray}
  F_\mathrm{e}(E,t) &=& 2\, A\,f_{\rm 0, e} \biggl({E\over
  \mathrm{m_\mathrm{p}}c^2}\biggr)^{-\gamma} v_\mathrm{shock}
  \tau_\mathrm{loss}\frac{1}{\gamma-1} \nonumber \\ & &
  \left\{1-(1-t/\tau_\mathrm{loss})^{\gamma-1} \right\} \hspace{1.cm} \text{for
  $ t< \tau_\mathrm{loss}$}
\label{int_dist_f}
\end{eqnarray}
and 
\begin{equation}
F_\mathrm{e}(E,t) = 2\, A\,f_{\rm 0,e} \biggl({E\over
   \mathrm{m_\mathrm{p}}c^2}\biggr)^{-\gamma} v_\mathrm{shock}
   \tau_\mathrm{loss}\frac{1}{\gamma-1} \quad \text{for $ t>\tau_\mathrm{loss}
   $} \nonumber \\
\end{equation}
   
Finally, we obtain the synchrotron spectrum by convolving
$f_\mathrm{e}(E,x,t)$, respectively $F_\mathrm{e}(E,t)$, with the synchrotron
emission spectrum of a single electron.

We will compare the synchrotron emission, predicted from the spatially
integrated distribution function (eq.\ref{int_dist_f}), to the observed radio
emission from the bridge.  The synchrotron emission derived from the local
distribution (eq.\ref{dist_f}) will be used to calculate the spectral index
$\alpha$ between 1.49 (respectively 1.40 for UGC 813/6) and 4.89 GHz at $x=0$
which will be compared to the observed value at the center of the bridge.

\subsection{Choice of the parameters}

Apart from the energy available for particle acceleration, estimated in
Sect. 3.1, we need to specify several additional parameters in order to
calculate the value of the expected radio emission from the bridge.  These
parameters are (i) the magnetic field strength, (ii) the energy density of the
radiation field, both determining the radiative energy losses, and (iii) $T$,
the time elapsed since the beginning of the interaction.  The adopted
parameters are listed in Tab.~\ref{data2}.  In the following we describe how
they were estimated and the uncertainty associated with them.

Most of the parameters are, at least roughly, constrained by the observations.
The magnetic field can be estimated from the minimum energy requirement.
Condon et al. (1993) derived the average magnetic field strength
with this method to be about 7~$\mu$G in the bridge of UGC~12914/5.  Condon et
al. (2002) derived values between 5~$\mu$G (south) and 9~$\mu$G (north) for the
bridge in UGC~813/6. We adopt their value for UGC 12914/5 and take an
intermediate value of $B = 7 \mu$G for UGC~813/6.

The radiation energy density in the bridges is low since practically no star
formation is taking place.  We can estimate the energy density due to star
formation in a similar way as done by Condon et al. (2002), from the observed
fluxes in the blue and FIR.  From the blue magnitudes of UGC~813 and UGC~816
($B_0$ = 13.7 and 13.2, respectively, from NED), a flux of $2.5\times 10^{-13}$
Wm$^{-2}$ follows. Adding the FIR flux of $1.7\times 10^{-13}$ W m$^{-2}$ we
obtain a total flux of $4.2\times 10^{-13}$ Wm$^{-2}$. From this, we estimate
a photon energy density in the bridge of about 0.3 eV cm$^{-3}$.  For
UGC~12914/5 we derive with $B_0$ = 12.51 and 13.06 for UGC~12914 and UGC~12915,
respectively, (from NED), a flux of $4.6\times 10^{-13}$ Wm$^{-2}$ in the
blue, and, together with a FIR flux of $3.7\times 10^{-13}$ Wm$^{-2}$, a total
flux of $8.3\times 10^{-13}$ Wm$^{-2}$.  From this, we estimate  the energy
density of the radiation field in the bridge to be about 0.5 eV cm$^{-3}$.
To these values we have to add the energy density due to the Cosmic
Microwave Background of 0.26 eV cm$^{-3}$ (e.g. Longair 1997).

Condon et al. (1993, 2002) estimated the time elapsed since the beginning of
the collision from the distance and relative velocity of the systems. They find
$T \approx 3 \times 10^7$ yr (for H = 0.75) for UGC 12914/5 (Condon et
al. 1993) and $T \approx 4 \times 10^7$ yr for UGC 813/6 (Condon et
al. 2002). The latter value would be slightly lower ($T \approx 3 \times 10^7$)
taking into account the different collisional velocity derived in Sect. 3.1.
We adopt values  in these ranges (see Tab.~\ref{data2}).  These values
are in addition constrained by a fit to the observed spectral index in the
center of the bridge.

From the energy density of the magnetic field and the radiation field, the
radiative energy loss time scale $\tau_\mathrm{loss}$ is derived.  The spectral
index in the center, $\alpha-\mathrm{predicted}$, depends in a very sensitive way on
$T/\tau_\mathrm{loss}$, if $T/\tau_\mathrm{loss}\approx 1$.  E.g. for the
values of $\urad$ and $\ub$ as in Tab. 2 for UGC 12914/5 we derive for $T = 2.5
\, 10^7$ yr ( respectiveley $2.8 \, 10^7$ yr, $3.0 \, 10^7$ yr) values of
$\alpha-\mathrm{predicted}$ of 1.3 (respectively 1.4, 1.6).  We can therefore use this dependence
as a further constraint on the combination of parameters $\ub$, $\urad$ and $T$
by comparing our model predictions to the observed value of $\alpha$ in the
center.  Whereas the influence of $T/\tau_\mathrm{loss}$ on $\alpha$ is rather
strong, the effect on the predicted total synchrotron emission is only minor.
E.g. for the values of $T$ as in the example above ($T = 2.5 \, 10^7$ yr, $2.8
\, 10^7$ yr, $3.0 \, 10^7$ yr) we estimate a total synchrotron emission
$P_{\nu}$-predicted, high = $1.8\, 10^{22}$ W Hz$^{-1} $, $1.7\, 10^{22}$ W
Hz$^{-1}$, $1.65\, 10^{22}$ W Hz$^{-1}$.

Thus, even though we calculate the spectral index in a simplified model,
(e.g. neglecting diffusion), which introduces some uncertainties when comparing
it to the data, it still provides a useful observational constraint on the
combination of $\ub$, $\urad$ and $T$. This constraint, together with the fact
that each parameter is reasonably well constrained by individual observations,
makes the uncertainty introduced by these parameters small. Overall, the
predicted radio emission at 1.49 GHz (respectively 1.40 GHz for UGC 813/6)
depends mainly on the available energy in the collision, whereas the other
parameters have only a secondary influence on the result.

\subsection{Results for the radio emission}

In Tab.~\ref{data2} we list the predicted values for the synchrotron emission.
We list a range of values which is given by the parameters that provide the
main uncertainty in our estimate: (i) The uncertainty in the molecular gas mass
which is mostly due to the uncertainty in the conversion factor from CO to
H$_2$. The gas mass directly determines the total kinetic energy. We adopt a
Galactic conversion factor as an upper limit, as
well as a 4 times lower factor, as suggested by Braine et
al. (2003) and Zhu et al. (2007).  (ii) The uncertainty in the efficiency of
particle acceleration in shocks.  We adopt a range of realistic values between
10\% -- 30\% .  We also list the predicted value for the spectral index,
$\alpha-\mathrm{predicted}$ between 1.49 GHZ and 4.89 Ghz (for UGC~12914/5), and between
 1.40 GHz and 4.89 GHz (for UGC~813/6), 
  respectively, in the center of the bridge.

In addition, for UGC 813/6 the uncertainty in the relative velocity (and thus
in the total available kinetic energy) is relatively high because the total
mass of the system, necessary for a dynamical study and an estimate of the
relative velocity, could not be determined from the HI line profile and had to
be estimated from the HI mass (Condon et al. 2003).  

The given range of predictions for the radio emission from the bridges
encompasses the observed values in both cases. Given the uncertainties, which
lie mainly in the energy that goes into the acceleration of relativistic
electrons, the agreement is satisfactory. The model calculations thus show that
acceleration by interstellar shocks caused by the galaxy-galaxy interaction is
indeed able to explain the radio emission from the bridge.

\section{High-energy gamma-ray emission}

Although not the main topic of this paper, it is clear that the interaction of
galaxies considered here will also lead to the acceleration of gamma-ray
producing very high-energy particles, both nuclei and electrons, in the form of
the distribution $f_\mathrm{acc}$, cf. Eq. (7). The visibility of the
acceleration process also in high-energy \grs would be an independent argument
for the model presented. Although this is not likely for present-day \gr
-instruments -- essentially due to the relatively large distance of the objects
-- it is therefore worthwhile, as a corollary, to investigate their expected
\gr flux. In the present context only a short estimate will be given, whereas
details are deferred to a seperate paper.

Concentrating on UGC 12914 for the following, $E_\mathrm{kin} = (0.49 \mbox{ to
} 1.4)\times 10^{58}$~erg, cf. Table 2. When the shocks have gone through the
interacting ISM of the two galaxies, which is the case at about the present
epoch, this energy is in the form of thermal and nonthermal particle energy. A
fraction $\Theta = 0.1 \mbox{ to } 0.3$ of this is by assumption in nonthermal
energy in relativistic particles, predominantly nuclei. This corresponds to
$E_\mathrm{acc} \sim 10^{57}$~erg which is roughly $10^7$ times more energy
than available from a single supernova remnant which liberates $10^{51}$~erg of
hydrodynamical explosion energy.

The bridge volume $V$ is estimated as $V \approx \pi R^3$, where $R \approx
10$~kpc \citep{condon93}. Then, using the masses for atomic and molecular hydrogen
in the bridge (cf. Tab. 2),  the gas density comes out to be $0.05 \mbox{ to
} 0.14$~cm$^{-3}$, which is rather small.

An analytical estimate for the integral hadronic \gr emission, from
$\pi^0$-production by collisions of energetic protons with gas nuclei and
subsequent decay into two \grs, is given in Eq.~(9) of \citet{drury94} for gamma
energies $E$ large compared to 100 MeV:
\begin{eqnarray}
F( > E) & \approx&  9 \times 10^{-11} \Theta \left(\frac{E}{1 \mathrm{TeV}}\right)^{-1.1}
\left(\frac{E_\mathrm{SN}}{10^{51}\mathrm{erg}}\right) \left(\frac{\mathrm{d}}{1
\mathrm{kpc}}\right)^{-2} \nonumber \\ 
& &\left(\frac {n}{1 \mathrm{cm}^{-3}}\right )
\mathrm{photons}~ \mathrm{cm}^{-2}~ \mathrm{s}^{-1}
\label{gamma_general}
\end{eqnarray}

Inserting the values $E_\mathrm{kin} = 10^{58}$~erg, $d=61 \mathrm{Mpc}$, and
$n=0.05 \mbox{ to } 0.14$~cm$^{-3}$ results in 
\begin{eqnarray}
  F( > E) & \approx& 9 \times 10^{-11} \Theta \left(\frac{E}{1 \mathrm{TeV}}\right)^{-1.1}
  (1.35 \mbox{ to } 3.6)\times 10^{-4} \nonumber \\
& & \mathrm{photons}~ \mathrm{cm}^{-2}~ \mathrm{s}^{-1}
\label{gamma_specific}
\end{eqnarray}

The last factor $(1.35 \mbox{ to } 3.6)\times 10^{-4}$ is very small compared
to unity, where a comfortable detection with ground-based imaging Cherenkov
telescopes is to be expected \citep{drury94}. On the other hand, the lowest \gr
flux from an astrophysical source detected until now was 
{  
$F( >220~\mathrm{GeV}) = 5.5 \times 10^{-13} \, \mathrm{photons}~ \mathrm{cm}^{-2}~
\mathrm{s}^{-1} $. }
This measurement was done with the H.E.S.S. telescope system
for the nearby starburst galaxy NGC253 \citep{acero09}. Taking this result as a
yard stick, the expected hadronic flux from UGC 12914 at \gr energies above {  220
GeV} is still {  a factor of 3-9} below this minimum flux, even for
the most optimistic values of the parameters discussed above. It can be shown
that the expected Inverse Compton flux is still by two orders of magnitude
lower, essentially as a result of the age $T\approx 3 \times 10^7$~yrs which
already leads to radiative cooling of the radio synchrotron electrons.

The situation would be more favorable for the future \gr instrument {\it
  Cherenkov Telescope Array (CTA)} whose sensitivity is expected to be a factor
of ten higher than that of the H.E.S.S. telescope system. For {\it CTA} UGC
12914 might therefore be marginally detectable.

As already mentioned above, the reason for the low \gr flux from both UGC
12914, and UGC 813/6 as well, is their comparatively large distance. Indeed if,
for example, UGC 12914 was at the distance of NGC 253, which is between 2.6 Mpc
and 3.9 Mpc, then its flux would be by a factor between 289 and 625
higher. This would be comfortably detectable even by present Northern
Hemisphere ground-based \gr instruments like {\it VERITAS} and {\it
  MAGIC}. Similar conclusions hold at GeV energies for the LAT instrument on
{\it Fermi}.

The present estimate indicates that such interacting systems are not very
promising sources of high-energy \grs, unless they are located rather close-by,
essentially just beyond the Local Group. It is therefore interesting to note
that radio observations may be able to tell us more about the general
nonthermal processes in such systems than \gr astronomy can do at the
present time.

\section{Concluding remarks}

From the above results it is concluded that the radio synchrotron emission from
the bridge region between the two galaxies can be explained by
the acceleration of a new population of relativistic electrons at the
large-scale shock waves that necessarily result from the supersonic collision
of the respective interstellar media. This process is not related to star
formation and therefore neither to the heating of dust by UV photons from
massive stars. As a consequence, there is no reason to
expect the standard radio/FIR correlation to hold for such a system. The
observations of these bridge systems and the fact that they show an excess of
radio emission with respect to the standard radio/FIR correlation can therefore
not be used as an argument against a  calorimeter model for
this correlation.

The process discussed in this paper may have broader implications for the
production of energetic particles, often summarily called cosmic rays, in
interacting astrophysical systems. An example may be the production of gas
fragments in galaxy clusters, when gas-rich galaxies enter the intracluster
medium and get partly stripped of their gas content by tidal interactions or by
more direct, hydrodynamic interactions with existing cluster galaxies up to
incomplete mergers \citep[e.g.][]{voexu94}. In all such cases the flow
velocities can be expected to be supersonic for
the relatively cold interstellar gas of the participating galaxies, resulting
in shear flows as well as supersonic compressions. 
  This is not
  necessarily the same process that happens when subclusters merge in the
  formation of larger clusters, because the diffuse gas involved is usually
  quite hot already in such cases, with the consequence that the particle
  acceleration process is rather inefficient. The process corresponds much more
  closely to the rather rare gravitational accretion of cold gas onto massive
  clusters.
 
  A similar process has been discussed more recently for Stephan's Quintett, a
  group of galaxies at a comparable distance of about 80 Mpc, by
  \citet{xu03}. In the intragroup medium of this system, a large shock ridge
  is detected which shows pronounced radio continuum and x-ray emission. The
  shock is produced by the collision of ISM from two galaxy group members.  The
  radio continuum has a spectral index of 0.93 \citep{xu03} which shows that it
  comes from moderately-aged synchrotron emission. Most likely, the origin of
  this radio emission is due to acceleration of relativistic electron in the
  shock in a very similar way as outlined here for the bridge systems.

Altogether, {   the two interacting galaxy pairs, discussed here, are possibly
examples of what might have happened much more frequently at early stages of structure
formation}, when primordial galaxies had already developed magnetic fields as a
consequence of early star formation, but when they were still likely to interact
strongly with neighboring structures of a similar character. 

\begin{acknowledgements}
UL acknowledges financial support from the research project 
AYA2007-67625-C02-02 from the Spanish Ministerio de Ciencia y
Educaci\'on and from the Junta de Andaluc\'\i a.
\end{acknowledgements}

\end{document}